\documentclass[pra,aps,twocolumn,footinbib,floatfix,showpacs]{revtex4}

\usepackage{times,amsmath}
\usepackage{epsfig}
\usepackage{amsmath}
\usepackage{amsfonts}
\usepackage{amssymb}
\usepackage{graphicx}%
\providecommand{\U}[1]{\protect\rule{.1in}{.1in}}

\newcommand{\TMYAG}{Tm$^{3+}$:YAG }

\begin{document}
\title{Efficiency optimization for Atomic Frequency Comb storage}

\author{M. Bonarota, J. Ruggiero, J.-L. Le~Gou\"et, T. Chaneli\`ere}
\email{thierry.chaneliere@lac.u-psud.fr}
\affiliation{Laboratoire Aim\'e Cotton, CNRS-UPR 3321, Univ. Paris-Sud, B\^at. 505, 91405 Orsay cedex, France}

\pacs{42.50.Md, 42.50.–p, 42.50.Gy, 03.67.-a}

\begin{abstract}
We study the efficiency of the Atomic Frequency Comb  storage protocol. We show that for a given optical depth, the preparation procedure can be optimize to significantly improve the retrieval. Our prediction is well supported by the experimental implementation of the protocol in a \TMYAG crystal. We observe a net gain in efficiency from 10\% to 17\% by applying the optimized preparation procedure. In the perspective of high bandwidth storage, we investigate the protocol under different magnetic fields. We analyze the effect of the Zeeman and superhyperfine interaction.
\end{abstract}
\maketitle

\section{Introduction}
\label{intro}

The prospect of a quantum memory (QM) for light naturally emerges in the atomic physics community since it deals with the most fundamental interaction of light with matter. Atomic vapors have specifically drawn the attention because they are model systems with good coherence properties. Experimental results rapidly follow with convincing proof-of-principle realizations. Most of them are based on stopped-light experiment and derived from the Electromagnetically Induced Transparency phenomenon. More generally, the protocols exploit a two-photon transition to benefit from a long live time in the ground state (spin coherence) \cite{polzikRMP}. A complete storage requires a large optical depth which also limits the signal bandwidth. As a consequence, the storage bandwidth is usually narrow (megahertz range in practice). The quest of wideband storage requires drastic changes. The protocols and the material should intrinsically offer a broad interaction linewidth. Inspired by impressive realizations in the classical domain \cite{mitsunaga1990248, lin1995demonstration}, the photon-echo techniques in rare-earth-ion doped crystals (REIC) appear as a prime alternative.

REIC have indeed also remarkable coherence properties at low temperature. The large natural inhomogeneous broadening gives a large potential storage bandwidth. It is not a limitation for the coherent transient phenomena. It has to be compared to the narrow homogeneous width (3-4 orders of magnitude narrower) to yield a high number of spectral channels \cite{Crozatier:06,Gorju:07}. This is alternatively interpreted as a high multi-mode capacity \cite{nunn}, crucial for quantum repeaters application.

Despite efficient demonstrations of classical storage \cite{carlson1983storage}, the photon-echo techniques should also be consider with precaution. A QM for light imposes different constraints. For example, the standard two-pulse photon echo (2PE) can be extremely efficient benefiting from intrinsic amplification (population inversion) of the medium. This regime is inappropriate for a QM since it deteriorates the fidelity  \cite{ruggiero:053851}. New specific protocols then emerge keeping the advantage of the technique (efficiency and bandwidth) but getting around the limitation of the 2PE (fidelity degradation): The Controlled Reversible Inhomogeneous Broadening (CRIB)  \cite{nilsson, alexander2006photon,kraus2006quantum} and Atomic Frequency Comb (AFC) \cite{AFCTh}. Theses techniques fully use the primary feature of the REIC (narrow homogeneous width, large inhomogeneous broadening, spectral hole-burning preparation procedures ...).

The CRIB protocol has been successfully implemented with record efficiencies \cite{hetet2008electro, tittel-photon}. It does not suffer from the previously mentioned limitations of the 2PE because the population mainly stays in the ground (no inversion). The CRIB has been implemented experimentally for weak light fields at the single photon level \cite{RiedmattenCRIB}. This recent experiment shows a low noise level given by technical limitations. In the same context, the AFC has been recently proposed \cite{AFCTh}. Experimental results followed immediately afterwards demonstrating also good noise feature and large efficiencies \cite{AFCexp, chaneliere2009efficient}. As compared to this realizations where the signal is stored directly in the optical coherence (optical AFC), the storage has also been performed in the spin coherences (Raman AFC) \cite{AFCPr}. This critical step opens the way for long memory time, on-demand retrieval and high multi-mode capacity.

The CRIB and the AFC have a lot in common since they are both inspired by the photon-echo technique. They both involve a rephasing of the optical coherences. The long storage time is obtained by a Raman transfer to the long-lived hyperfine coherences. The main difference comes from the preparation procedure. For the CRIB, a narrow absorption profile is tailored within the natural inhomogeneous broadening. An electric field gradient is then used to control and reverse the artificial broadening of a Stark effect sensitive medium. The hole-burning selection procedure has been optimized and may require a large number of independent pulses \cite{rippe2005experimental}. For the AFC, the absorption profile exhibits a periodic distribution of very absorbing peaks. The preparation procedure is based on spectral hole-burning towards a shelving state as in CRIB. A periodic optical pumping spectrum produces the corresponding modulation of the population. This is sufficient to produce an optical AFC echo without external control of the atomic transition frequency. On the one hand, the experimental implementation is not restricted to Stark sensitive REIC. On the other hand, a rudimentary preparation procedure based on the repetition of pulse pairs has allowed proof-of-principle demonstrations with relatively large efficiencies \cite{AFCexp, chaneliere2009efficient}. This two points explain the rapid experimental progresses \cite{AFCexp, chaneliere2009efficient, AFCPr} since the theoretical proposal of the AFC \cite{AFCTh}.

In this paper, we study the preparation procedure. We specifically optimize the pulse sequence to improve the efficiency of the optical AFC. We experimentally design the optical pumping spectrum to obtain the appropriate atomic comb. We also investigate the influence of the magnetic field. It changes the level structure and then the broadband optical pumping dynamics. This study is relevant in the large bandwidth regime required for quantum repeaters application.

The paper is arranged as follows. In section \ref{I}, we show how the preparation sequence can be optimized for an efficient optical AFC retrieval. The results are analyzed by modelizing the population dynamics during the preparation stage. In section \ref{II}, we study the efficiency under a varying magnetic field and shows the influence of the Zeeman and superhyperfine effects.

\section{Comb tailoring and storage efficiency}\label{I}

The generation of an echo out of a spectrally periodic medium is the key ingredient to explain the AFC storage. This is interpreted as a dipole rephasing in the time-domain \cite{AFCTh, chaneliere2009efficient} or alternatively as diffraction on a spectral grating \cite{chaneliere2009efficient, Sonajalg:94}. This latter approach is fruitful to calculate the retrieval efficiency. We then describe the ideal comb shape that lead to the maximum efficiency. This analysis is well supported the experimental results obtained in \TMYAG.

\subsection{Optimized AFC shape}

The key idea of the AFC protocol is the storage capacity of a spectrally periodic medium. This observation connects the AFC to the well-know three-pulse photon echo (3PE) or more generally to space-and-time-domain holography. For the 3PE, the first two engraving pulses create a population modulation. The third pulse is diffracted and produces an echo. The AFC is a direct descendant of the 3PE in that sense. To increase the population modulation depth by accumulation, the engraving sequence can be repeated. The AFC preparation procedures used so far also involve accumulated pulse pairs \cite{AFCexp, chaneliere2009efficient}. The AFC has also some major differences with the 3PE which makes it especially interesting for quantum storage. On one side, the protocol implies a Raman transfer to obtain a long memory time, on-demand retrieval and perfect efficiency. On the other side, the comb shape is a very singular periodic modulation. In the limit of perfect retrieval, narrow absorbing peaks should be separated by fully transparent regions \cite{AFCTh}. This situation requires a large initial optical depth. This is usually a limiting factor in practice. We now can see how to optimize the comb shape to obtain the largest efficiency for a given optical thickness.

\subsubsection{Retrieval efficiency}\label{reteff}

We here calculate the retrieval efficiency. In the spectral domain, it is deduced from the comb structure. We consider the diffraction of a spectral grating. This model has been explained previously, we here simply review the main results \cite{AFCexp, Sonajalg:94}.

We use the Bloch-Maxwell equations assuming the slowly varying amplitude and the rotating wave approximations \cite{allen1987ora}. In the weak signal limit \cite{crisp1970psa}, the field propagation and atomic evolution are given by

\begin{equation}\label{MB}
\begin{array}{ll}
\partial_z\Omega(z,t)+ \frac{1}{c}\partial_t\Omega(z,t)=
-\displaystyle\frac{i }{2\pi}\int_\Delta \alpha\left(\Delta\right) \mathcal{P}(\Delta;z,t) \\[0.4cm]
\partial_t\mathcal{P}(\Delta;z,t)=-\left(i \Delta + \gamma\right)\mathcal{P}(\Delta;z,t)
-i\Omega(z,t)
\end{array}
\end{equation}

where $\Omega(z,t)$ is the Rabi frequency and $\mathcal{P}(\Delta;z,t)=\mathrm{u}(\Delta;z,t)+i\mathrm{v}(\Delta;z,t)$ is the polarization including the in-phase and out-of-phase components of the Bloch vector ($\gamma$ is the homogeneous line-width). We model the inhomogeneous broadening by a frequency dependent absorption coefficient $\alpha\left(\Delta\right)$ including the coupling with individual dipoles and the inhomogeneous lineshape.

The spectral periodic structure produces a series of echos. We decompose the absorption coefficient distribution in a Fourier series $\displaystyle \alpha\left(\Delta\right)= \sum_n \alpha_n e^{-i n \Delta T}$ ($1/T$ is the comb period). Successive echos whose respective amplitudes are $a_p$ appear at times $pT$ \textit{i.e.} $\displaystyle \Omega(z,t)=\sum_{p\geq0} a_p\left(z\right) \Omega\left(0,t-pT\right)$, $p\geq0$ for causality reason.

We'd like to forewarn the reader. We here summarize the propagation through the spectrally prepared medium. For reason of simplicity, the comb is modeled by a frequency selective absorption coefficient. It may give the wrong impression that the retrieval is a pure absorption effect. The causality imposes that the successive echos appear after the arrival of the signal. This natural assumption is actually strong because it connects the field amplitude and phase by the Kramers-Kronig relation. The interplay between the field amplitude, phase and energy will be specifically addressed later in section \ref{energy}.

So far, the different amplitudes of successive echos are recursively calculated from the propagation equation in the Fourier space. We here focus on the amplitude of the first echo. It defines the retrieval efficiency $|a_1\left(L\right)|^2$ ($L$ is the crystal length) and is given by successive integration with the boundary conditions $a_0\left(0\right)=1$ matching the incoming field intensity and $a_1\left(0\right)=0$:

\begin{equation}\label{a0a1}
\begin{array}{ll}
\partial_z a_0\left(z\right) =\displaystyle -\frac{1}{2} \alpha_0  a_0\left(z\right) \\[0.3cm]
\partial_z a_1\left(z\right) =\displaystyle -\frac{1}{2} \left[\alpha_0  a_1\left(z\right) + 2 \alpha_{-1}  a_0\left(z\right) \right]
\end{array}
\end{equation}
We finally obtain an analytic expression for the efficiency connecting the comb shape and the first two coefficients of the absorption coefficient Fourier expansion.

\begin{equation}\label{etag0g1}
\eta\left( L\right)  =\displaystyle \left|\alpha_{-1} L \right|^2  e^{-\alpha_{0} L }
\end{equation}

It has been independently calculated by S{\~o}najalg \textsl{et al.} in the context of space-and-time-domain holography \cite[eq. (13)]{Sonajalg:94}. We now address the problem of optimizing the comb shape to maximize the efficiency.

\subsubsection{Shape optimization}\label{optimizedshape}

This question can be seen as a pure mathematical constrained optimization. Our analysis is based on different physical situations.
The efficiency (eq. \ref{etag0g1}) is a monotonic function of $\left|\alpha_{-1}\right|$ and $\alpha_{0}$. The largest efficiency should be obtained for a large $\left|\alpha_{-1}\right|$ and a negative $\alpha_{0}$. The first condition makes sense because the retrieval is directly generated by the modulation proportional to $\left|\alpha_{-1}\right|$. Next, a negative value $\alpha_{0}$ (gain) means that the medium is amplifying.

This situation is realistic. One can indeed imagine that the atomic population is modulated between the ground state (absorption) and the excited (population inversion). In that case, the absorption distribution is modulated between negative and positive values both bounded by $\alpha_{M}$ and $-\alpha_{M}$ respectively ($\alpha_{M}$ is limited by the available number of atoms). The efficiency can be larger than one. The presence of gain is interesting for classical storage by providing internal amplification. It has been investigated experimentally with REIC \cite{Crozatier:05, SjaardaCornish:00}.

The perspective of quantum storage is changing the constraints. As mentioned in the introduction, the presence of gain reduces the fidelity of a quantum memory \cite{ruggiero:053851}. Population inversion is then not allowed. This means that the absorption $\alpha\left(\Delta\right)$ is always positive and bounded by $\alpha_{M}$. This is the situation we consider from now.

Our analysis shows that classical and quantum storage impose different constrains. It tells us also what should be the shape of the comb to optimize the efficiency. Since we impose for $\alpha\left(\Delta\right)$ to be always positive and bounded by $\alpha_{M}$, one can guess that the constrains  are saturated: $\alpha\left(\Delta\right)$ reaches its boundaries. In other words, one expects that the absorption keeps its maximum accessible values \textsl{i.e.} $0$ or $\alpha_{M}$. The absorption profile should then have a square shape. The width of the square peaks is calculated by maximizing the efficiency (eq. \ref{etag0g1}).

\begin{figure}[th]
\includegraphics[width=8.5cm]{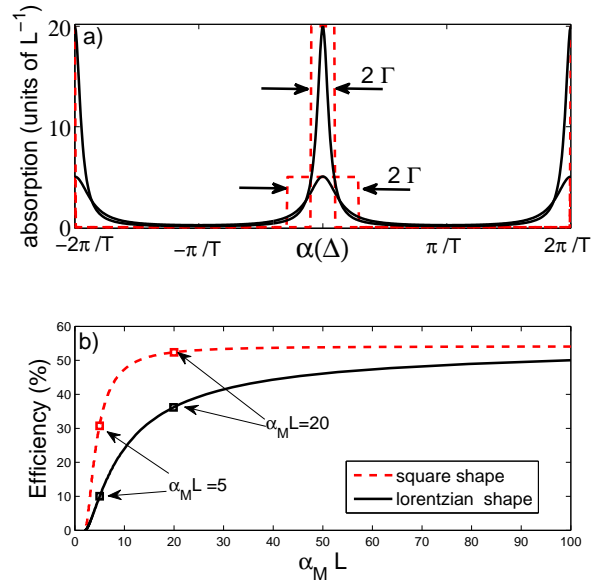}\caption{(color online) a. For a given maximum optical thickness ($\alpha_{M} L=5$ and $\alpha_{M} L=20$ in that case), we plot the combs that optimize the efficiency in the two following cases: the comb is made of well-separated lorentzian peaks (solid black line - see \cite{AFCTh} for details) or square-shaped peaks (red dashed line). b. Optimum efficiency for lorentzian  and square shape peaks as a function of the maximum optical depth $\alpha_{M} L$. The four square symbols give the efficiency for the shapes in fig. \ref{PlotCarreLorentzOPT_article}.a).}
\label{PlotCarreLorentzOPT_article}
\end{figure}

So we define $\alpha^S$, a $2\pi/T$-periodic function, between $-\pi/T$ and $\pi/T$ with $\alpha^S\left(\Delta\right)=\alpha_{M}$ between $-\Gamma$ and $\Gamma$ and $0$ elsewhere (see dashed lines in fig. \ref{PlotCarreLorentzOPT_article} a. for example).

The coefficients of the Fourier series are then $\alpha^S_{0}=\alpha_{M} \displaystyle \frac{\Gamma T}{ \pi}$ and $\alpha^S_{-1}=\alpha_{M} \displaystyle \frac{\sin\left(\Gamma T\right)}{\pi}$. $\Gamma$ is the half-width at half-maximum. The proper width $\Gamma_\mathrm{OPT}$ that maximizes the efficiency is

\begin{equation}\label{gamma-opt}
\Gamma_\mathrm{OPT}^S=\displaystyle  \frac{1}{T}\arctan\left(\frac{2 \pi}{\alpha_{M} L}\right)
\end{equation}

It only depends on the maximum optical thickness $\alpha_{M} L$. The corresponding efficiency is then given by
\begin{equation}\label{eta-opt}
\eta_\mathrm{OPT}^S= \left( \frac{\alpha_{M} L}{\pi}\right)^2 \sin^2\left(\Gamma_\mathrm{OPT}^S T\right)  e^{-2 \alpha_{M} L \Gamma_\mathrm{OPT}^S T/\pi}
\end{equation}

At low optical thickness, $\Gamma_\mathrm{OPT}^S=\pi/\left(2T\right)$ corresponding to a one-half duty cycle (or finesse equals two). The width should then decrease for very absorbing media. The same behavior has been derived by assuming that the comb is made of well-separated lorentzian peaks  \cite{AFCTh, chaneliere2009efficient}. Analytic formula are obtained as well for the optimum width maximizing the efficiency (see \cite{AFCTh} for details). We then compare this two situations (see fig. \ref{PlotCarreLorentzOPT_article}). 

We clearly observe an improvement for the square-shaped peaks (red dashed line in fig. \ref{PlotCarreLorentzOPT_article}.a). The efficiency of the optical AFC is limited to 54\% in the forward direction. This limitation is due to the absence of gain as discussed previously and is the same for the CRIB protocol \cite{kraus2006quantum}. This asymptotic behavior is due to the trade-off between a good absorption of the signal and a moderate re-absorption at the retrieval. We see that even a limited optical thickness (typically $\alpha_{M} L\sim 10$) gives an efficiency close to 50\% for the square shape. A comparable improvement is expected in the backward direction (Raman AFC).

\subsubsection{Field and energy propagation}\label{energy}

Our analysis connects the absorption spectrum and the retrieval mechanism. As previously mentioned, it may give an incorrect image of the protocol. This point is made explicit when the energy propagation is considered in the spectral domain. It is derived from the Bloch-Maxwell equations (eqs. \ref{MB}).

\begin{equation}\label{energy_propag}
\partial_z |\widetilde {\Omega}\left(z,\omega \right)|^2= - \alpha\left(\omega\right)\otimes \mathcal{L}\left(\Delta\right)  |\widetilde {\Omega}\left(z,\omega \right)|^2
\end{equation}

The optical power spectral density $|\widetilde {\Omega}\left(z,\omega \right)|^2$ is defined by the Fourier transform of field in pulsed regime. The  absorption is convoluted by the homogeneous lineshape $\mathcal{L}\left(\Delta\right)= \displaystyle \frac{\gamma}{\gamma^2+\Delta^2}$. The convolution is negligible as soon as the spectral structures are much larger than the homogeneous width $\gamma$. This is the case for the experiment. The equation tells us that the medium acts as a spectral filter. Nevertheless, it may be misleading to properly interpret the AFC protocol. This is particularly manifest for square-shaped absorption peaks. At large optical thickness where the highest efficiency is expected, the peaks should be extremely narrow. The power spectrum is then fully transmitted (eq. \ref{energy_propag}). In other words, the absorbing spectral filter has a marginal influence on the transmitted energy even if the protocol is very efficient in this regime as demonstrated in \ref{optimizedshape}. In this situation, the energy should mainly migrate from the signal to the first AFC echo (54\%). The rest goes to the successive echos. This happens without modification of the optical power spectrum. This property is essentially due to a frequency selective phase response through the propagation. As previously mentioned, the echo propagation obeys the causality. This assumption actually connects the field amplitude and phase by the Kramers-Kronig relation. We do not explicitly calculate the phase propagation but it is embedded in the model because of the causality.

Our discussion is also an interesting way to underline the singularity of a square-shaped combs. The transmission spectrum $e^ {-\alpha \left(\Delta \right) L}$ has also a square shape. The transmission peaks have the same width $\Gamma$, whatever the maximum optical depth. This is not the case for lorentzian absorption peaks. In transmission, the structures become wider for a large optical depth. They are typically broadened by a factor $\sqrt{\alpha_M L}$. The transparent dispersive regions are then truncated. This reduces the energy transmission whose importance has been emphasized previously. This physical argument \textit{a posteriori} justifies the expected improvement for square-shaped combs.

\subsection{Experimental square-shaped comb}
In order to support our prediction, we experimentally prepare square-shaped peaks in the absorption spectrum. We expect a significant improvement as compared to the rudimentary preparation procedure that has been used so far (accumulation of pulse pairs see \cite{AFCexp, chaneliere2009efficient}). This latter has the advantage of simplicity. It only needs two pulses of equal intensity. This produces a periodic spectrum for the optical pumping light. It is then not surprising that this procedure creates a comb like structure through the hole-burning process. The production of the comb has been alternatively interpreted as Ramsey fringes \cite{AFCexp}. The relationship between the pumping light spectrum and the final comb shape is not obvious and is related to the atomic population dynamics. We specifically address this question in section \ref{disc}. We nevertheless can guess intuitively that a square-shaped light spectrum with a $2\pi/T$ period generates a square-shaped comb through the optical pumping process. We now prepare this specific optical spectrum and use it for the preparation procedure.

\subsubsection{Square-shaped optical spectrum with a $2\pi/T$ period}

A $2\pi/T$-periodic spectrum for the pumping light $|\widetilde {\mathcal{E}}\left(\omega \right)|^2$ is produced by a pulse train in the time-domain. The temporal separation $T$ imposes the period and the pulse duration gives the total bandwidth. The fine structure of the spectrum depends on the relative pulse amplitudes.

Experimentally, the temporal shaping of the preparation pulses is realized by an acousto-optic modulator (AOM) connected to an Arbitrary Waveform Generator (AWG 520 Tektronik). We here use the same optical setup that have been previously described \cite{chaneliere2009efficient}. This allows a complete control in amplitude and phase of the preparation pulse sequence $\mathcal{E}\left(t \right)$. We'd like to obtain a a field spectrum $\widetilde {\mathcal{E}}\left(\omega \right)$  which has a constant value between $-\Gamma$ and $\Gamma$ and $0$ elsewhere in the interval $\left[-\pi/T;\pi/T\right]$. In the time-domain, it is a pulse train with relative amplitudes $P\left(k\right)=\mathrm{sinc}\left(k \Gamma T/2 \right)$ ($k$ is integer, $P\left(k\right)$ can be negative).

By controlling the AWG RF amplitude and phase sent to the AOM, we create an arbitrary sequence of pulses. The non-linear RF response of the AOM should be taken into account in order to faithfully control the field spectrum. The RF to optical intensity response is recorded and is used as a correction factor to obtain the $P\left(k\right)$ amplitudes of each pulse (for negative values of $P\left(k\right)$, the RF phase is flipped). We truncate the series to $-30\leq k\leq 30$ corresponding to 61 pulses. The pulse duration is $300$ns adapted to the AOM bandwidth and the temporal separation is $1.5\mu s$ ($1/T=666$ kHz) \cite{chaneliere2009efficient}. We experimentally verify the field spectrum by heterodyning the pulse sequence with a shifted reference from the same laser.

\begin{figure}[th]
\includegraphics[width=7cm]{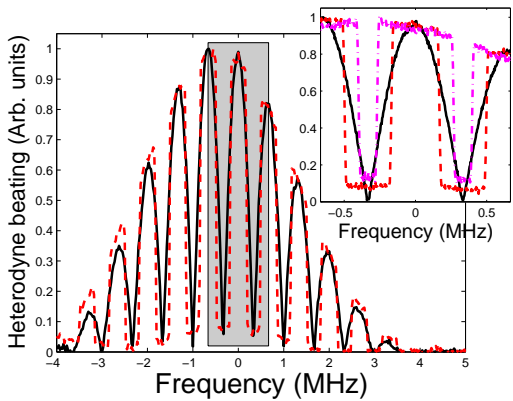}\caption{(color online) Heterodyne beating of the preparation pulse sequence for $\Gamma=\pi/\left(2T\right)$ with $1/T=666$ kHz (red dashed line). Since the reference is monochromatic, it directly gives the field spectrum  $|\widetilde {\mathcal{E}}\left(\omega \right)|$. We compare it with the spectrum of a pulse pair $P\left(0\right)=P\left(1\right)=1$ (solid black line). Inset: Central part of the spectrum $\left[-1/T;1/T\right]$. We add the pulse sequence corresponding to $\Gamma=\pi/\left(5T\right)$ that is also used in the experiment (magenta dashed dotted line).}
\label{TraitHeterodyne_article}
\end{figure}

We observe in fig. \ref{TraitHeterodyne_article} a good overall control of the field spectrum $\widetilde {\mathcal{E}}\left(\omega \right)$. By applying 61 pulses, we obtain a square-shaped spectrum with a $2\pi/T$ period. The peak width is changed by modifying the pulse sequence (see inset).

The spectrum is asymmetric about the carrier frequency. This is explained by an asymmetry in the AOM frequency response. It strongly depends on the alignment of the device itself and the optical fiber placed behind. More importantly, we note an imperfect contrast in the modulated spectrum. This is especially true for its central part. It can be due to a relative distortion of the pulses. This point is not surprising because of the previously mentioned non-linear response of the AOM. The potential effect of the pumping dynamics will be discussed later (see section \ref{disc}). We prepare four different time sequences. The first is a pulse pair (PP-sequence) and is taken as reference. The three others S$_{1/2}$, S$_{1/3}$ and S$_{1/5}$-sequences have a square-shaped spectrum with different widths respectively $\pi/\left(2T\right)$, $\pi/\left(3T\right)$ and $\pi/\left(5T\right)$. We now experimentally evaluate their influence on the AFC efficiency.

\subsubsection{Measured AFC efficiencies}\label{measAFC}
We implement the protocol in 0.5\% doped-\TMYAG crystal \cite{chaneliere2009efficient}. A 210G magnetic field is applied along the [001] crystalline axis and splits the ground and excited levels into a nuclear spin doublet by $\Delta_g=6 $MHZ and $\Delta_e=1.3$ MHz respectively. The comb structure is obtained by optical pumping between these long-lived spin levels. The laser polarization is also applied along the [001] axis. The crystal is immersed in liquid helium at 2.3K. This ensures a narrow homogeneous linewidth ($\gamma\simeq2\pi\times 5$kHz) and then an accurate spectral tailoring.

The four previously described pulse sequences (PP, S$_{1/2}$, S$_{1/3}$ and S$_{1/5}$) represents an elementary pattern whose total duration is $100 \mu s$. This pattern is repeated 5000 times to accumulate the population in the shelving state with a low pumping power (weak area pulses). The resulting comb depends on the pulse sequence and the pumping power. It is probed after 50ms waiting time for a complete decay of the excited state population. 

\begin{figure}[th]
\includegraphics[width=8.5cm]{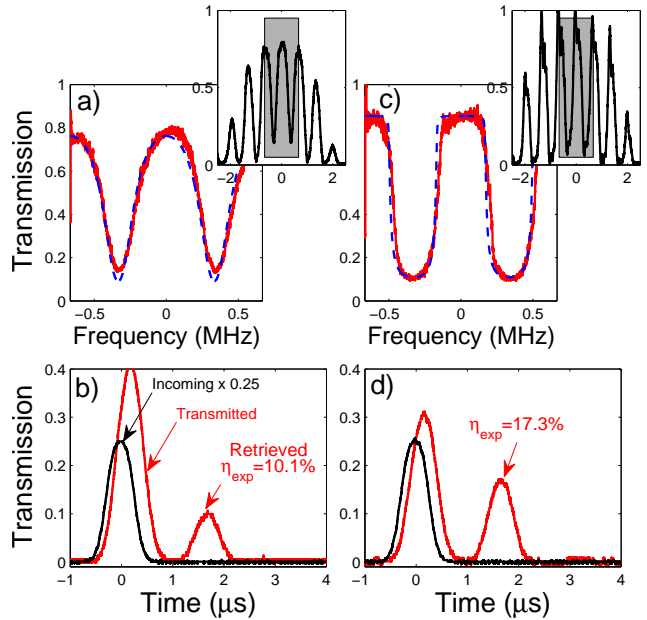}\caption{(color online) a. and c. Central part of the absorption spectrum after the PP and S$_{1/2}$ sequences respectively. We fit the spectra by a simple optical pumping model (dashed blue line, see section \ref{disc} for details). The insets show the same spectrum on a larger bandwidth. b) Optical AFC echo of a weak $450$ns signal pulse after the PP-sequence (corresponding to a). d. Optical AFC after the S$_{1/2}$-sequence (corresponding to d). The pumping power is adjusted independently for both sequences to optimize the retrieval efficiency.}
\label{TraitMaxEffafc1_dentr5_article}
\end{figure}

The independent weak probe beam is controlled by a second AOM for frequency scanning and temporal shaping. First, it is used to measure the comb shape by sweeping the AOM carrier frequency. We obtain the absorption spectrum (see fig \ref{TraitMaxEffafc1_dentr5_article} a. and c.). Next, a $450$ns signal pulse is sent into the medium. It is longer than the preparation pulses to ensure a good spectral overlap of the signal bandwidth and the comb. This probe pulse is absorbed and retrieved at a later time $T=1.5\mu s$ (see fig \ref{TraitMaxEffafc1_dentr5_article} b and d).

\begin{figure}[th]
\includegraphics[width=8.5cm]{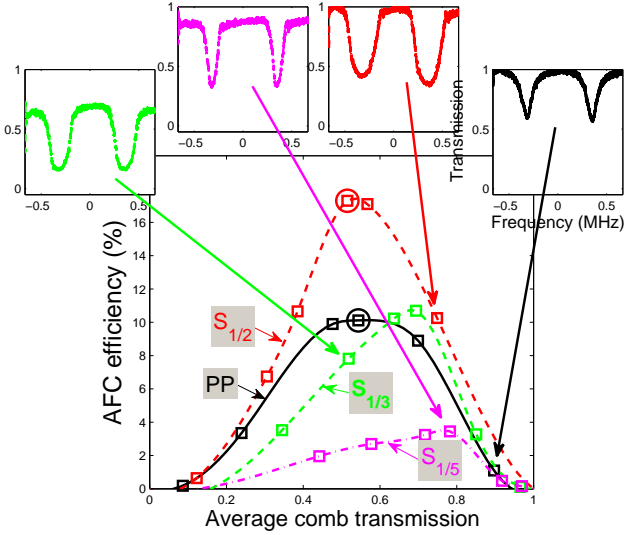}\caption{(color online) Efficiency comparison between the PP, S$_{1/2}$, $_{1/3}$ and S$_{1/5}$ sequences. The lines connecting the measurements are used to guide the eye. The four insets show an absorption spectrum for each sequence. The corresponding measurements are indicated by arrows. The two circles represent the measured efficiencies and spectra that have been analyzed in fig \ref{TraitMaxEffafc1_dentr5_article}.}
\label{ComparSeq_article}
\end{figure}

We see a net gain between the PP and the S$_{1/2}$-sequence since they give an efficiency of 10.1\% and 17.3\% respectively. This is expected since the two sequences generate different comb shape (see fig \ref{TraitMaxEffafc1_dentr5_article} a and c). The sharp edges of absorption spectrum approaching a square shaped structure for the S$_{1/2}$-sequence (see fig \ref{TraitMaxEffafc1_dentr5_article} c) explains a gain in efficiency. There are also noticeable differences between the field spectrum and atomic comb (for example the inset of fig. \ref {TraitHeterodyne_article} and the fig. \ref{TraitMaxEffafc1_dentr5_article}c can be compared). First, the absorption spectrum is visibly smoothed out by the optical pumping dynamics. Moreover, we cannot obtain a fully contrasted comb. More precisely, the transmission should be 1 at the maximum and few percent at the minimum limited by the sample optical depth ($\alpha_{M} L\sim 4-5$ in our case). The maxima and minima of the absorption spectrum strongly depend on the pumping power. The latter is chosen to optimize the retrieval efficiency for the PP and S$_{1/2}$ sequences. In order to have a good understanding of the preparation population dynamics, we vary the pumping power and record the corresponding efficiency (see fig. \ref{ComparSeq_article}). We decide to use the mean transmission value (central part of the absorption spectrum) as abscissa to plot the different efficiencies. It as the advantage to be easily accessible experimentally. It is also obtained from the model that we develop in section \ref{disc}.

We expect different results for different square widths as previously discussed in section \ref{optimizedshape}. We indeed see that the square width should be properly chosen to obtain the highest efficiency (eq. \ref{gamma-opt}). We then also use the sequences S$_{1/3}$ and S$_{1/5}$ to observe this dependency. We recognize the features derived from our analysis of the optimum comb shape (section \ref{optimizedshape}) \textit{i)} a square like shape of the comb improves significantly the retrieval. \textit{ii)} the efficiency strongly depends on the square width of the optical spectrum.

We should be able to predict the efficiency from the atomic comb shape (eq. \ref{etag0g1}). This procedure demands a fine knowledge of the absorption spectrum \textit{i.e.} the average value $\alpha_0$ and the modulation $\alpha_{-1}$. We measure both coefficients by recording the transmission spectrum. The results of eq. \ref{etag0g1} can then be compared to the independently measured efficiencies. This approach is unfortunately unsuccessful to fit the data set presented in fig. \ref{ComparSeq_article}. To explain this discordance, we see two reasons. The first one is technical. It is not easy to accurately measure a large optical thickness. In that case, a weak transmission can be biased by experimental artefacts \cite{chaneliere2009efficient}. The second one is fundamental. We implicitly assume in section \ref{reteff} that the atomic comb is uniform along the crystal length. This allows a complete analytical calculation. In the contrary case, $\alpha_0$ and $\alpha_{-1}$ (in eqs. \ref{a0a1}) would be z-depend. We can question this assumption. The comb is produced by a specific pulse preparation sequence. During the propagation, this sequence can be distorded by the atomic comb under construction. As a consequence, the resulting comb would not be uniform along the crystal. In order to analyze the impact of these two effects and have a better physical understanding of the AFC limitation, we develop a simple pumping model to discuss the previous data. 

\subsubsection{Discussion - optical pumping dynamics}\label{disc}
The optical pumping dynamics is a key element to have at least a qualitative understanding of the current limitations. We here briefly develop a model to account for the frequency selective hole-burning preparation.

The rate equation limit of the Block-Maxwell equations have been successful to predict the hole shape in high resolution spectroscopy with monochromatic pumping \cite{allen1987ora}. This approach is perfectly valid in our case since each sequence pattern satisfy the weak area approximation. In that case, each sequence (duration $T_p=100 \mu s$) induces an incremental small population change $w \rightarrow w + \delta w$ that is given by $\delta w =\displaystyle \frac{1}{2 \pi} \mathcal{L}\left(\Delta\right) \otimes |\widetilde {\mathcal{E}}\left(\Delta \right)|^2 w$. The optical spectrum  $|\widetilde {\mathcal{E}}\left(\Delta \right)|^2$ ($\widetilde {\mathcal{E}}$ is the Fourier transform of the pulse sequence) is convoluted by the homogeneous lineshape $\mathcal{L}\left(\Delta\right)= \displaystyle \frac{\gamma}{\gamma^2+\Delta^2}$. The population difference between the excited and the ground state is $w=n_1-n_2$.

\begin{figure}[th]
\includegraphics[width=7.5cm]{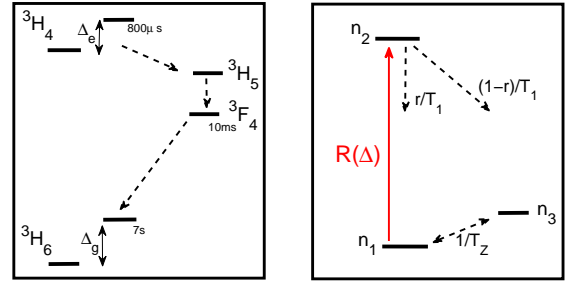}\caption{Left: Thulium level structure. Right: Equivalent three-level system for optical pumping model.}
\label{LevelStruct}
\end{figure}

Since we repeat each sequence to accumulate the population in a shelving state, we model the preparation by a continuous frequency selective pumping rate $R\left(\Delta\right)=\displaystyle \frac{\delta w}{w T_p}$

\begin{equation}\label{rate}
R\left(\Delta\right) =\displaystyle \frac{1}{2 \pi T_p} \mathcal{L}\left(\Delta\right) \otimes |\widetilde {\mathcal{E}}\left(\Delta \right)|^2 
\end{equation}
The hole-burning process is due to at least one shelving state. In our case, even if most of the population ends up in the long-lived Zeeman state, the pumping dynamics also involve the $^{3}$F$_{4}$ bottle-neck state (lifetime 10ms). To simplify the problem, we model our system as an equivalent three level system (see fig. \ref{LevelStruct}).

The rate equations for the populations $n_1$,$n_2$, and $n_3$ (see fig. \ref{LevelStruct}) are then
\begin{equation}\label{rateN}
\begin{array}{lll}
\partial_t n_1\left(\Delta\right) = \frac{1}{2}R\left(\Delta\right) \left(n_2-n_1\right) + \frac{r}{T_1} n_2 -  \frac{1}{T_Z} \left(n_1-n_3\right)\\ [.4cm]
\partial_t n_2\left(\Delta\right) =  \frac{1}{2} R\left(\Delta\right) \left(n_1-n_2\right) - \frac{1}{T_1} n_2\\ [.4cm]
\partial_t n_3\left(\Delta\right) =  \frac{1-r}{T_1}n_2 - \frac{1}{T_Z}\left(n_3-n_1\right)
\end{array}
\end{equation}
At the end of the pumping sequence, the populations reach a steady state $n_1^S$,$n_2^S$, and $n_3^S$. Before sending the signal, we wait for a complete decay from the excited state. The populations should now be $n_1^P=n_1^S+r n_2^S$,$n_2^P=0$, and $n_3^P=n_3^S+\left(1-r\right) n_2^S$. The absorption spectrum on the 1-2 transition is given by
\begin{equation}\label{alpha3niv}
\alpha\left(\Delta\right) =\displaystyle \alpha_M  n_1^P\left(\Delta\right) = \alpha_M \frac{\left(1+r\right) R \left(\Delta\right) T_1 + 2}{R \left( \Delta \right) T_1 /\epsilon + 4}
\end{equation}
where $\epsilon=\displaystyle \frac{1}{\left(1-r\right)T_Z/T_1+3}$. The parameter $\epsilon$ should be much smaller than one. The population then accumulates is 3 because $\left(1-r\right)/T_1>>1/T_Z$.

This formula connects the optical spectrum and the absorption profile. To obtain an analytical solution for the AFC efficiency, we have to assume that the optical spectrum is uniform over the entire length of the medium. In that case only, the atomic comb is uniform as well. That is an underlying assumption of the efficiency formula (eq. \ref{etag0g1}). This supposition will be discussed later on.

Such a model allows us first to predict the absorption spectrum for any predefined pumping sequence. We here focus on the PP and S$_{1/2}$-sequences which yield the highest efficiencies. The PP-sequence produces a cosine-modulated spectrum. In our case, since $1/\gamma >> T$, we obtain for the pumping rate at the center of the spectrum (eq. \ref{rate}): $R\left(\Delta \right) \propto \displaystyle 1+ \exp\left(\gamma T \right) \cos\left(\Delta T \right)$. By using the formula \ref{alpha3niv}, we fit the absorption spectra produced by the PP-sequence with different pumping powers. We see a typical comparison between our model (dashed blue line in fig \ref {TraitMaxEffafc1_dentr5_article} a) and the experiment.

The S$_{1/2}$-sequence has been designed on purpose to generate a square-shaped optical spectrum. The convolution by a lorentzian smooths out the edges and the pumping rate now reads as $R\left(\Delta\right) \propto \displaystyle 2\sum_{p} \arctan \left(\frac{2\pi\left(p+1/2\right)-\Delta T}{\gamma T}\right)$. We also fit the experimental absorption spectrum in fig \ref {TraitMaxEffafc1_dentr5_article} c).

The agreement is satisfying for both sequences. Our model is able to properly predict the observed transmission spectrum.

We now interpret the connection between the heterodyne beat and the absorption spectrum. We have seen that the S$_{1/2}$-sequence induces a residual background reducing the contrast of the modulated spectrum (see fig. \ref{TraitHeterodyne_article}). Our model does not include this component but properly reproduces the observed absorption spectrum. We conclude that this background does not have a major impact. It seems that the reduction of the comb contrast through the optical pumping dynamics is mainly due to the homogeneous linewidth (convolution effect). This limitation is fundamental and not technical. 

\begin{figure}[th]
\includegraphics[width=8cm]{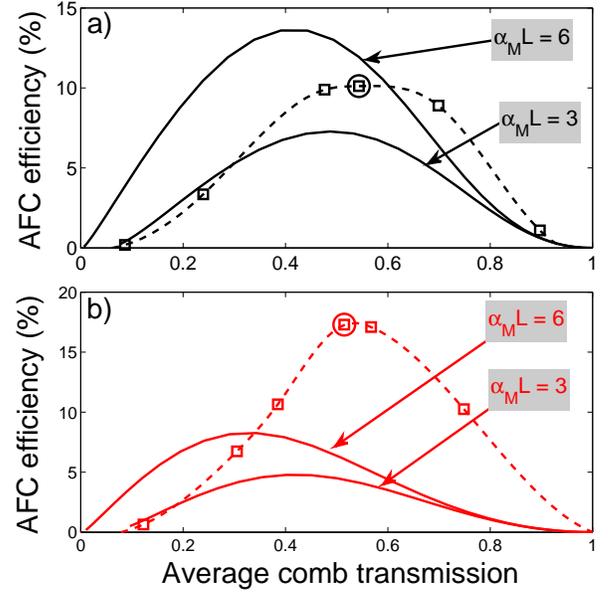}\caption{(color online) Comparison between the measured (same as in fig. \ref{ComparSeq_article}) and the predicted efficiencies for the PP (a) and S$_{1/2}$-sequences (b). For each sequence, we plot the prediction of our pumping model with $\alpha_{M} L=3$ $\alpha_{M} L=6$ for the PP (a - solid black lines) and S$_{1/2}$-sequences (b - solid red lines).}
\label{ComparPPS12_model_article}
\end{figure}

The comparison with the observed absorption spectrum validates our approach of modelling the pumping process. Combining eqs. \ref{alpha3niv} and \ref{etag0g1}, we are able to predict the observed efficiency (in fig. \ref{ComparSeq_article}) for the PP and S$_{1/2}$-sequences. As we do not independently measure the pumping power which gives the amplitude of $\widetilde {\mathcal{E}}$, we also use the mean transmission value (central part of the absorption spectrum). The average transmission is indeed predicted by our model (eq. \ref{alpha3niv}) and is used in abscissa. We then plot the efficiency for different pumping power (see fig. \ref{ComparPPS12_model_article}). The variable pumping power parametrizes the curve. The only remaining adjusting parameter is the initial optical depth $\alpha_{M} L$.

The initial absorption is typically $\alpha_{M} L\sim 4-5$. It is not completely easy to accurately measure a large optical depth. The efficiency strongly depends on $\alpha_{M} L$. To illustrate this dependency and consider a larger range, we plot the expected efficiencies for $\alpha_{M} L=3$ and $\alpha_{M} L=6$ (solid black lines and red dashed lines in fig. \ref{ComparPPS12_model_article} a and b for the PP and S$_{1/2}$-sequences respectively).

The general qualitative behavior is properly reproduced. Without pumping light, the mean transmission is $e^{-\alpha_{M} L}$ and the efficiency zero (the comb is absent). At very high pumping power, the medium is completely transparent and there is no AFC echo neither. The power broadening effect is indeed smoothing out the periodic modulation of the pumping spectrum and uniformly bleaching the sample.
Quantitatively we observe a major discrepancy between the curves. We cannot reproduce the position of the maximum even if we consider optical depths ($\alpha_{M} L=3$ and $6$) which are largely surrounding the experimental value. This may sound surprising since we fit accurately the measured transmission spectrum (see fig \ref {TraitMaxEffafc1_dentr5_article} a and c). To explain this discordance, we go back to the conclusion of section \ref{measAFC}. If the atomic comb was uniform over the entire length of the medium, we should be able to fit both the measured transmission spectrum and experimental efficiencies. Our optical pumping model predicts independently both and should allow us to get around the experimental artefacts biasing the transmission measurements. We then expect that the atomic comb is actually non-uniform. This would certainly strongly modify the propagation of the AFC echo (in eqs. \ref{a0a1}, $\alpha_0$ and $\alpha_{-1}$ would be z-depend) and change the final efficiency. But it sounds possible to fit properly the finally observed transmission spectrum. It indeed results from an average absorption spectrum over the propagation but could be fitted by an equivalent uniform profile. In that sense, the AFC efficiency would be more sensitive to the uniformity of the comb than the transmission spectrum.

The propagation of the pumping spectrum through the sample produces certainly a non-uniform atomic comb. To account for this propagation effect, one can include the back-action of the absorption spectrum (propagation of the pumping light). This leads to two coupled equations that can be solve numerically. This work is currently under investigation in our group.

\section{Influence of the magnetic field}\label{II}
The previous analysis is based on an optical pumping model. We simplify the level structure (see fig. \ref{LevelStruct}) to obtain an equivalent structure and stay close to the physical intuition. If we neglect the influence of the $^{3}$F$_{4}$ bottle-neck state, such a simplification seems to be appropriate as soon as the ground and excited state Zeeman splittings ($\Delta_g$ and $\Delta_e$) are much larger than the comb bandwidth. The shelving state is then out of resonance. Spectral hole-burning (SHB) is an alternative way to consider the pumping dynamics. SHB has been widely used as a fine spectroscopic tool with monochromatic laser especially for Zeeman interaction and superhyperfine interaction \cite{liu2005spectroscopic}. Under magnetic field, the spectrum is usually enriched by side-structures as holes or anti-holes. Following this analogy, the preparation of a comb by optical pumping should also lead to side combs and anti-combs. If this structures are completely out off resonance, one could indeed treat the preparation as a population accumulation in a single far-off-resonance shelving state. For a weak magnetic field or when a large bandwidth is demanded, the side structures overlap the main comb. They can be caused by the Zeeman (hyperfine) or the superhyperfine effect. We specifically study these situations here.

\subsection{Zeeman effect}\label{zeeman}

\begin{figure}[th]
\includegraphics[width=8cm]{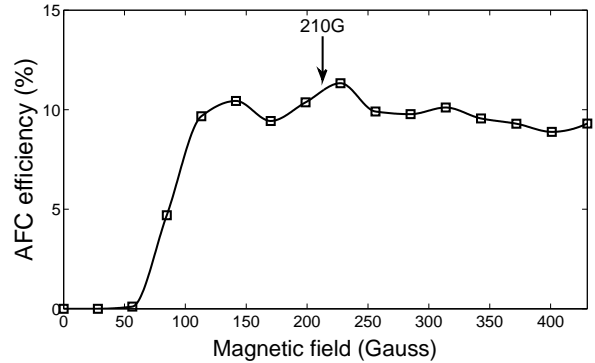}\caption{AFC efficiency as a function of the magnetic field (the solid line is used to guide the eye). The error bar are smaller that the marker size. The 210G value that has been used for the previous experiments is indicated by an arrow (see text for details).}
\label{Trait_vsB_article}
\end{figure}

The magnetic field is applied along the [001] crystalline axis. In that case, all the excited sites are equivalent. The Zeeman splittings have been measured previously: $\delta_g=28$ kHz/G and $\delta_e=6.0$ kHz/G respectively for the ground and the excited state \cite{louchet:035131}. The experimental 210G magnetic field produces shifts of $\Delta_g=5.9$ MHz and $\Delta_e=1.3$ MHz. The AFC bandwidth is typically 4MHz limited by the 300 ns preparation pulse duration (see inset of fig. \ref{TraitMaxEffafc1_dentr5_article} c. for example). Even if the ground state splitting is larger than the bandwidth, this not the case for the excited state. This modifies the pumping dynamics. In SHB spectroscopy a monochromatic laser creates a hole at the central frequency. In the present situation with four levels, we observe also two side holes positioned at $\pm \Delta_e$, four anti-holes at $\pm \Delta_g$ and $\pm \left(\Delta_g-\Delta_e\right)$. Our preparation procedure is then creating a comb centered at the central frequency, two side combs at $\pm \Delta_e$ and four anti-combs at $\pm \Delta_g$ and $\pm \left(\Delta_g-\Delta_e\right)$. Since $\Delta_g>>\Delta_e$, the influence of the anti-combs should be negligible for a 210G magnetic field. Nevertheless we cannot neglect the two side combs at $\pm \Delta_e = \pm 1.3$ MHz as it is smaller than the bandwidth. To minimize the interplay between these structures, we have intentionally chosen the magnetic field in order to have $\Delta_e\simeq 2/T$. This \textit{a priori} assumption deserves further study. To question it, we use the PP-sequence. We then simply measure the optical AFC efficiency as a function of the applied magnetic field (see fig. \ref{Trait_vsB_article}).

We observe a strong influence of the magnetic field on the efficiency. We analyze this figure as follows.

At low magnetic field ($<$50G), the efficiency is zero. This is directly due to the population lifetime in the Zeeman state. We cannot observe long-lived holes by SHB spectroscopy. The preparation is then inefficient. We observe an echo only if the population lifetime is significant ($>$100G in our case).

The range 100-140G is very singular since $\Delta_g$ and $\Delta_e$ are both smaller than the bandwidth ($\simeq$ 4MHz). The situation is complex because the main comb, the side combs and anti-combs overlap each other. But nevertheless it leads to a high efficiency. If the side combs are shifted by the main comb spacing ($\Delta_e=p/T$, $p$ is integer), and the anti-comb by half of it ($\Delta_g=\left(p\prime+1/2\right)/T$, $p\prime$ is integer), the resulting profile is very contrasted. In other words, the preparation spectrum is optically pumping and de-pumping at the same time within the four-level system. The two mechanisms act accordingly to increase the comb contrast. We cannot satisfy all the conditions experimentally when $\Delta_g$ and $\Delta_e$ are kept smaller than the bandwidth. A 130G magnetic field represents the best trade-off between the conditions: $\Delta_e=1.2/T$ and $\Delta_g=5.5/T$. We indeed observe a local maximum of the efficiency for this value (see fig. \ref{Trait_vsB_article}). It supports our analysis.

In the range 140-430G, only $\Delta_e$ is smaller than the bandwidth. An overlap of the side combs with the main one is possible. The matching is realized for 220G ($\Delta_e=2/T$) and 330G ($\Delta_e=3/T$). We also observe a local maximum of the efficiency at 220G close the value chosen for the previous experiments. For higher values of the magnetic field, the overlap is decreasing and the efficiency constant. 

Our study is particularly relevant for the optical AFC. We clearly see that high efficiencies are obtained when the four-levels are all contained in the interaction bandwidth. Such a configuration is less suitable for the Raman AFC. First, a spin doublet in the ground state is not sufficient to provide one shelving state plus two spin states (long coherence lifetime). Next, the Raman transfer demands to address independently the different transition. A overlap may not be appropriate especially for the ground state where the coherences are stored. A partial overlap in the excited state may be tolerated for the Raman AFC but this possibility has not been investigated so far.

\subsection{Superhyperfine effect}

\begin{figure}[th]
\includegraphics[width=8.5cm]{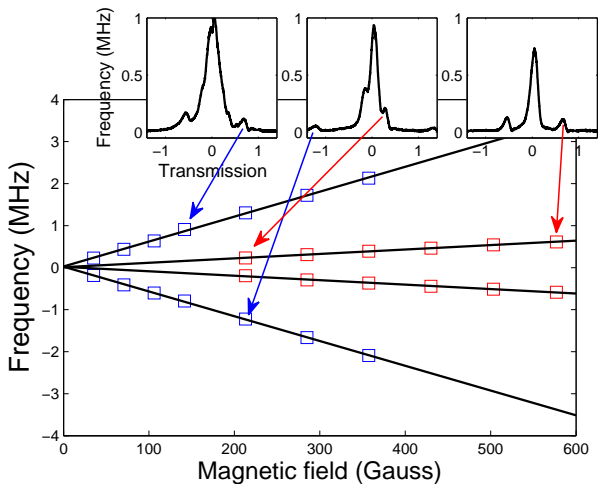}\caption{(color online) Spectral hole-burning spectroscopy of the Zeeman excited state splitting (blue markers) and the superhyperfine interaction (red markers). We plot the side holes position as a function of the magnetic field. Insets: Examples of the hole-burning spectra (for 110, 210 and 580G from left to right). We confirm previous measurements for the excited state Zeeman shift $\delta_e=5.91 \pm 1\%$ kHz/G \cite{louchet:035131}. For large magnetic fields ($>$210G), we see adjacent side holes that we attribute to the superhyperfine interaction with Al nuclei. The splitting is $\delta_S=1.05 \pm 1\%$ kHz/G.}
\label{Plot_Trait_SuperH_article}
\end{figure}

The superhyperfine effect corresponds to the next order of magnetic field induced interaction. It is produced by the surrounding nuclear spins on the dopant rare-earth \cite{Macfarlane:81}. It is usually studied with EPR or NMR spectroscopy. An optical detection makes the observation easier for luminescent centers \cite[p.72]{liu2005spectroscopic}. SHB spectroscopy can also be used to study this ion-ligand interaction \cite{PhysRevB.38.11061}. The superhyperfine effect is then indicated by the presence of adjacent anti-holes on the broader hyperfine or Zeeman structure. We perform SHB measurements for a varying magnetic field along the [001] crystalline axis (see fig. \ref{Plot_Trait_SuperH_article}).

We initially observe side holes at $\pm \Delta_e$ due to the Zeeman interaction (see fig. \ref{Plot_Trait_SuperH_article}). The previously mentioned anti-holes at $\pm \Delta_g$ and $\pm \left(\Delta_g-\Delta_e\right)$ are present but rapidly out of the spectrum range and small because of the large optical depth. The strongest surrounding nuclear moments are the aluminium ones. In that case, the population distribution is usually governed by $\Delta m_I=\pm 1$ spin flips  \cite{kaplianskii1987spectroscopy}. Our result $\delta_S=1.05 \pm 1\%$ kHz/G is consistent with the Al gyromagnetic ratio 1.11 kHz/G. One can be surprised by the intensity of the superhyperfine effect (hole depth) which is comparable to the Zeeman one. Nevertheless the Al ions occupy different sites in the crystal leading to different coupling strengths. Since we do not observe any quadrupole structure, we cannot distinguish them \cite{priv:superhyperfine}. Such an analysis demands an accurate spectroscopic study.

The superhyperfine interaction modifies the AFC preparation dynamics and generating side structures as the Zeeman effect does at lower magnetic field. At moderate magnetic field, the shift is smaller than the AFC peak and then broaden it. This reduces the contrast of the comb and then the efficiency. In our case, the peak width is typically 200kHz corresponding to a 180G superhyperfine shift. For a larger magnetic field, side structures should clearly appear. Even if because of the high number of side structures (Zeeman and superhyperfine), the matching between them is relatively unlikely for a given magnetic field.

The effect should be stronger for Kramers ions in high spin concentration matrices. Its influence on the AFC storage has already been observed \cite{AFCexp}. A systemic material study should be pursued especially because a high bandwidth and then a large magnetic field is desired for quantum repeaters application.

\section{Conclusion}
We carefully study the efficiency of the optical AFC protocol. We show that the preparation procedure can be significantly improved. The most efficient atomic comb have a square shape whose width should properly adjusted for a given optical depth. This requires an accurate control of preparation pulse sequence in amplitude and phase. We implement this technique in a \TMYAG crystal and increase the efficiency from 10.1\% to 17.3\%. The experimental results are consistent with our expectations. We develop a simple rate equation model to describe the optical pumping dynamics. Even if it properly reproduces the comb spectrum, we are unable to predict \textit{ab initio} the observed efficiencies. This discrepancy tells us that the propagation of the preparation pulses inside the medium and the backaction of the prepared atomic comb should be taken into account. This route is under investigation in our group.

We also study the effect of the magnetic field. This is relevant in the perspective of high bandwidth application. The Zeeman and the superhyperfine interactions have a strong influence on the optical AFC protocol. They can be beneficial or detrimental. We clarify these conditions.

Our work should be placed in the context of quantum storage where a Raman conversion has to be implemented (Raman AFC). Our analysis mainly focuses on the preparation stage of optical AFC but can be extended to understand the efficiency of the complete protocol.

\section*{Acknowledgments}
We thank C. Thiel and Y.C. Sun for their analysis of the superhyperfine effect. We'd like to thank D. Comparat and A. Fioretti for technical assistance on magnetic coils design and realization.

\end{document}